\documentclass[aps,pra,twocolumn,showpacs,showkeys,reprint]{revtex4-1}
\usepackage{graphics}
\usepackage{graphicx}
\usepackage{amsmath}

\begin{document}

\title{Mott lobes of the $S=1$ Bose-Hubbard model with three-body interactions}
\author{A. F. Hincapie-F}
\author{R. Franco}
\author{J. Silva-Valencia}
\email{jsilvav@unal.edu.co}
\affiliation{Departamento de F\'{\i}sica, Universidad Nacional de Colombia, A. A. 5997 Bogot\'a, Colombia.}

\date{\today}

\begin{abstract}
Using the density matrix renormalization group method, we studied the ground state of the one-dimensional $S=1$ Bose-Hubbard model with local 
three-body interactions, which can be a superfluid or a Mott insulator state. We drew the phase diagram of this model for both ferromagnetic and 
antiferromagnetic interaction. Regardless of the sign of the spin-dependent coupling, we obtained that the Mott lobes area 
decreases as the spin-dependent strength increases, which means that the even-odd asymmetry of the two-body antiferromagnetic chain is absent for local 
three-body interactions. For antiferromagnetic coupling, we found that the density drives first-order superfluid-Mott insulator transitions for even 
and odd lobes. Ferromagnetic Mott insulator and superfluid states were obtained with a ferromagnetic coupling, and a tendency to a ``long-range'' 
order was observed.
\end{abstract}


\maketitle

\section{Introduction}

The use of the optical lattices as an emulator for condensed matter physics has opened the door to new developments with ultracold 
atoms~\cite{Anderson-S98}, and the observation of a quantum phase transition from a superfluid to a Mott insulator state by 
Greiner {\it et al.}~\cite{Greiner-N02a} using $^{87}$Rb atoms increased the study of atoms confined in optical traps. This transition was predicted 
by Jaksch {\it et al.}~\cite{Jaksch-PRL98} considering the spinless Bose-Hubbard Hamiltonian, which has been widely studied, and its phase 
diagram is well known~\cite{Kuhner-PRB98,Kuhner-PRB00,Ejima-EL11}.\par 
The subsequent development of experimental techniques has allowed confining atoms in purely optical traps such that the degree of freedom of 
spin is not frozen, allowing the creation of spinor condensates, which exhibit new quantum physical properties, and this has stimulated the study of 
quantum magnetism with cold atom setups~\cite{StamperKurn-PRL98}. Alkali atoms such as $^{87}$Rb or $^{23}$Na have a nuclear spin $I=3/2$ and an 
electronic spin $J=\pm 1/2$, so that these atoms can be confined with a hyperfine spin state 1 or 2. The effective Hamiltonian to describe the 
behavior of spin-1 bosons in optical lattices was proposed by Imambekov {\it et al.}~\cite{Demler-PRL02,Imambekov-PRA03}. In order to obtain it, they considered two channels in the contact 
interaction. Therefore, the Hamiltonian contains a kinetic term and two local interaction terms, the two-body repulsion and a spin-dependent 
interaction. This model has been widely studied in one dimension by using mean-field approximations and numerical 
techniques, and its phase diagram is known~\cite{Imambekov-PRA03,Tsuchiya-PRA04,Rizzi-PRL05,Apaja-PRA06,Pai-PRB08,Batrouni-PRL09,Natu-PRB15}. There exist Mott 
insulator lobes and a superfluid region, among which quantum phase transitions occur. It has been shown that when the local spin interaction is 
antiferromagnetic, the system has an asymmetry in the Mott lobes: the even ones grow with an increase in the spin interaction parameter, while odd lobes 
decrease~\cite{Demler-PRL02}. Also, within the odd lobes a competition between a dimerized and a nematic singlet 
phase has been explored~\cite{Yip-PRL03,Rizzi-PRL05,Apaja-PRA06,Bergkvist-PRA06,Toga-JPSJ12}. On the other hand, for ferromagnetic interaction the lobes 
always decrease as the spin interaction parameter increases, and the magnetic ordering within the lobes is ``long-range''  ferromagnetic 
order~\cite{Batrouni-PRL09}. Furthermore, it has been shown that for antiferromagnetic interaction the phase transition is first-order for even lobes 
due to the coexistence of singlets in the two phases, while for odd ones it is a second-order transition~\cite{Batrouni-PRL09}.\par  
Several authors have studied this model in two-dimensional lattices, showing that the system in the superfluid phase exhibits a nematic order and the 
phase transition from superfluid to Mott insulator is first or second order if the density at the Mott lobe is even or odd, 
respectively~\cite{Toga-JPSJ12, Forges-PRB13,Forges-PRL14}.\par 
Recently, upon studying the dynamics of spin-1 bosons loaded in a 3D optical lattice, Mahmud {\it et al.} showed that the Hamiltonian to describe the 
physics of the system does not contain only two-body interactions: multi-body interactions make the model more accurate. Also, by 
means of the perturbation theory they derived an effective spin-dependent and spin-independent three-body interaction terms~\cite{Mahmud-PRA13}.\par  
For spinless bosons systems, local and non-local multi-body interaction terms have been considered during the last decade. For instance, B\"uchler 
{\it et al.}~\cite{Buchler-NP07} proposed that polar molecules in optical lattices can be tuned to a regime where three-body interactions play 
the dominant role, and Jhonson {\it et al.}~\cite{Jhonson-NJP09} obtained that two-body interactions generate effective higher-body interactions. 
Also, in 2010 Will {\it et al.}~\cite{Will-N10} showed experimental evidence of multi-body interactions using an interferometric technique for 
$^{87}$Rb atoms confined in an optical lattice, and after that, resonances were associated with higher-order processes in photon-assisted tunneling 
experiments~\cite{Ma-PRL11}. Other authors have stated that when considering complex geometries like the double-well optical lattice, the 
effective Hamiltonian contains strong three-body interactions tuned by lattice parameters~\cite{Paul-PRA15}. On the other hand, Daley 
{\it et al.}~\cite{Daley-PRA14} proposed an experimental setup using laser-assisted tunneling in a bosonic gas in which the three-body interactions 
dominate the physics of the system. Very recently, using Feshbach resonance, a novel way to get rid of the two-body interactions has been proposed. This 
interaction strength can be obtained as a function of the scattering length and an effective range. These two can be tuned until the two-body 
interaction term vanishes while three atoms interact at the same lattice site via three-body contact repulsion~\cite{Paul-PRA16}.\par 
Phase diagrams of spinless bosons under three-body interactions in one dimension have been studied and found by several authors.  For instance, 
bosons under on-site interactions~\cite{Chen-PRA08,JSV-EPJB12,Sowinski-PRA12,Varma-PRB14}, nearest neighbor interaction~\cite{Capogrosso-PRB09}, and contact three 
body interaction~\cite{SafaviNaini-PRL12} have been considered. An interesting result obtained for the on-site case was that the area of the 
first Mott lobe (two bosons per site) decreases while the second Mott lobe grows as the three-body strength increases.\par 
Taking into account that the physics of spinor bosons in one dimension is very interesting and the fact that multi-body interactions can generate 
new phases and the perturbative calculation of the three-body local interaction terms between spin-1 bosons, in the present paper we explore a 
one-dimensional system of spinor bosons that interact under local three-body terms. To our knowledge, this problem has not been studied, and we 
calculate the ground state energy and the correlation functions by using the density matrix renormalization group method for ferromagnetic and 
antiferromagnetic interactions. For both kinds of interactions, we found that the phase diagram has Mott insulator lobes and a superfluid region. We 
obtained that the area of the Mott insulator lobes decreases with the spin-dependent term, and also a quantum phase transition in the superfluid 
region was found.\par 
The structure of the paper is as follows: In Sec. \ref{sec2} we explain the Hamiltonian of the $S=1$ Bose-Hubbard model with three-body interaction. In 
Sec. \ref{sec3}  we show the evolution of the chemical potential and the correlation functions as a function of the system parameters for 
antiferromagnetic and ferromagnetic spin-dependent interaction; finding diverse phase diagrams for both cases. The last section, Sec. \ref{sec4}, 
contains a summary and conclusions.\par 

\section{\label{sec2} Model}
 
In 2013, Mahmud and Tiesinga made a perturbative calculation and found the effective expressions for the interaction among three spin-1 bosons. These  
expressions are local and are given in terms of the density and spin operators~\cite{Mahmud-PRA13}. The Hamiltonian of the unexplored problem of 
one-dimensional spinor bosons interacting through {\it only} three-body terms is given by
\begin{eqnarray}\label{Hgen}
\mathcal{H}=&-&t \sum_{i,\sigma} \left( b^{\dagger}_{i,\sigma} b_{i+1,\sigma}+ 
\text{h.c} \right) \nonumber \\ &+&\frac{V_0}{6}\sum_i \hat{n}_i(\hat{n}_i-1)(\hat{n}_i-2)\nonumber \\ 
&+& \frac{V_2}{6}\sum_i(\mathbf{\hat{S}}_i^2-2\hat{n}_i)(\hat{n}_i-2),
\end{eqnarray}
\noindent where  $b^{\dagger}_{i,\sigma}(b_{i,\sigma})$ creates (annihilates) a boson with spin 
component $\sigma$ on site $i$ of a one-dimensional optical lattice of size $L$, 
$\hat{n}_i=\sum_{\sigma}b^{\dagger}_{i,\sigma}b_{i,\sigma}$ and 
$\mathbf{\hat{S}}_i=\sum_{\sigma,\sigma '}b^{\dagger}_{i,\sigma}\mathbf{T}_{\sigma,\sigma '} b_{i,\sigma '}$ are the total number of particles and
 the spin operators on site $i$, respectively. $\mathbf{T}_{\sigma,\sigma '}$ are the spin-1 Pauli matrices.\par 
The first term in the Hamiltonian (\ref{Hgen}) is the kinetic energy modulated by $t$, which is a measure 
of the tunneling force between nearest sites. The second term represents the short-range interaction 
between three o more atoms at the same site, where $V_0$ is the on-site repulsion parameter. 
The last term in the Hamiltonian is the local spin exchange interaction, tuned by the $V_2$ parameter. The latter can be positive or 
negative, according to the nature of the interaction, ferromagnetic ($^{87}$Rb) or antiferromagnetic ($^{23}$Na), respectively. 
For a harmonic potential with frequency $\omega_f$, $V_0$ is given by $V_0=-1.34\frac{U_0^2}{(\hbar\omega_f)}$ 
and $V_2$ is given by $V_2=2\frac{U_2}{U_0}V_0$~\cite{Jhonson-NJP09}, where $U_0$ and $U_2$ are the two-body 
strengths parameters given by $U_0= 4\pi \hbar^2(a_0+2a_2)/3M$ and $U_2= 4\pi\hbar^2(a_2-a_0)/3M$, 
where $a_s$ are the scattering lengths for $S=0$ and $S=2$ channels, and  $M$ is the mass of the 
atom~\cite{Stamper-RMP13}. The fraction $\frac{U_2}{U_0}$ can be determined from experimental scattering 
lengths, and it is $0.04$ for $^{23}$Na and $-0.01$ for  $^{87}$Rb~\cite{Ho-PRL98,Ohmi-JPSJ98}.\par 
\begin{figure}[t]
\includegraphics[width=20pc]{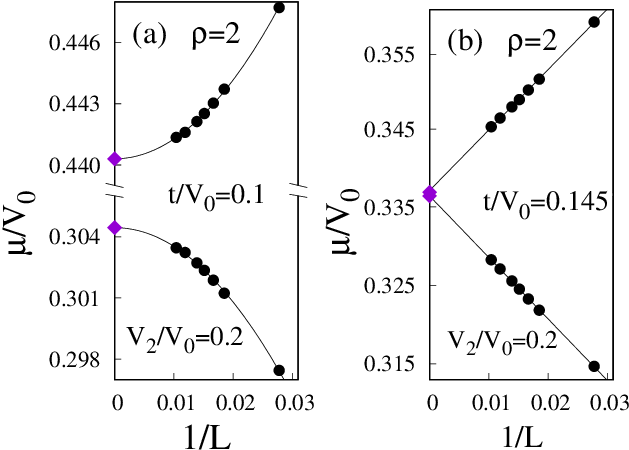}
\caption{\label{fig1} System size dependence on the chemical potential at $\rho =2$ and $V_2/V_0 =0.2$. The left 
panel shows a gapped state with an integer number of particles per site for $t/V_0=0.1$, and in the right panel the system is in a gapless 
state for  $t/V_0=0.145 $. The upper set of data corresponds to the energy for adding a particle, and the lower one to 
the energy for removing one. The values at the thermodynamic limit (diamonds) were obtained by fitting the DMRG results (circles) in the left (right) panel to the function 
$\mu^{p,h}(L)=\tilde{\mu}^{p,h}+ c_1/L + c_2/L^2$  ($\mu^{p,h}(L)=\tilde{\mu}^{p,h}+ c_3/L$), where $c_i$ are constants. The lines are visual guides}
\end{figure}
Note that for spinless bosons interacting through three-body on-site terms, the phase diagram exhibits Mott insulator regions, and a superfluid phase
surrounding them. However, in order to obtain a Mott insulator state, we need at least two bosons per site~\cite{JSV-EPJB12,Sowinski-PRA12}. When we 
consider spinor bosons, it is expected that insulator and superfluid phases will appear, and the border that separates these phases can be 
estimated by means of the energy gap for adding or removing one particle from the system, taking into account the canonical ensemble. This gap is given 
by $\Delta(L)=\mu^p(L)-\mu^h(L)$ where the chemical potential for adding or removing a particle is given by 
$$\mu^p(L)= E_0(L,S_{tot}^z,N+1) - E_0(L,S_{tot}^z,N),$$ 
 $$\mu^h(L)= E_0(L,S_{tot}^z,N) - E_0(L,S_{tot}^z,N-1).$$ 
\noindent $E_0(L,S_{tot}^z,N)$ being the ground state energy for $L$ sites, $N$ particles with $S_{tot}^z$ the $z-$component of the total spin.\par 
\begin{figure}[t]
\begin{minipage}{17.5pc}
\includegraphics[width=17.5pc]{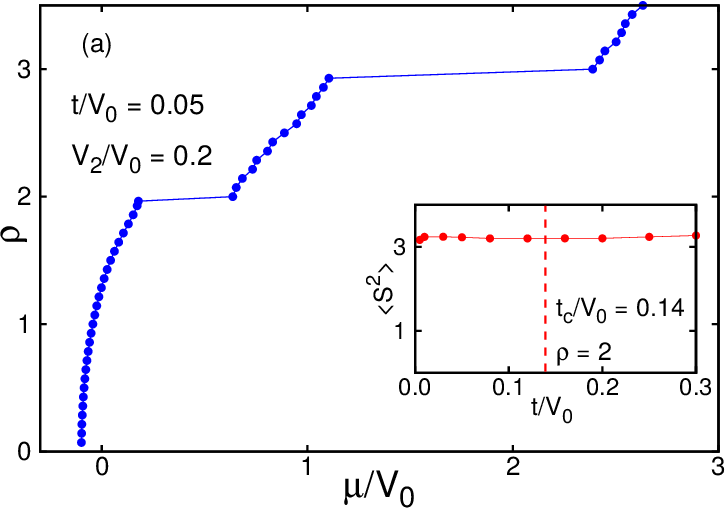}
\end{minipage}
\hspace{2pc}%
\begin{minipage}{17.5pc}
\includegraphics[width=17.5pc]{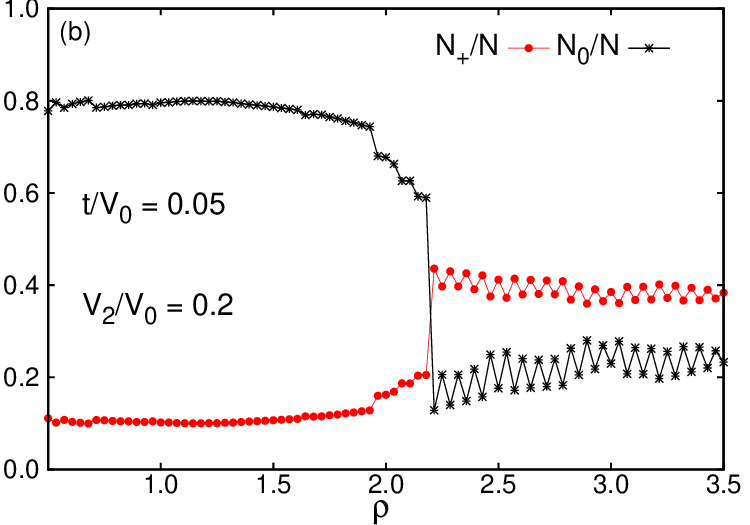}
\caption{\label{fig2}   
(a) The global particle density $\rho$ as a function of the chemical potential $\mu$ exhibits a Mott insulator plateau at integer densities greater 
than one. The inset shows the evolution of $<S^2>$ as a function of the hopping parameter $t/V_0$ for a density $\rho=2$. The vertical dashed line 
indicates the position of the Mott insulator-superfluid transition. (b) The spin population 
fractions $N_+/N$ and $N_0/N$ as a function of the global density $\rho$ for $t/V_0=0.05$, $V_2/V_0=0.2$, and $L=28$.}
\end{minipage}
\hspace{2pc}%
\end{figure}

\section{\label{sec3}  Results}
In order to determine the phase diagram of the Hamiltonian (\ref{Hgen}), we employed the density 
matrix renormalization group (DMRG) method with open boundary conditions \cite{White-PRL92}. We used the 
finite-size algorithm for sizes up to $L=96$ unless otherwise written. Fixing the parameters of our problem, we calculated the energy for 
different values of the total spin along $z$ ($S_{tot}^z=0,1 \text{ and }2$) and we obtained that the ground state is degenerate, which suggests 
that the Katsura theorems are valid for this model, and  therefore all calculations presented in this paper are in the 
sector with spin component $S_{tot}^z=0$ in order to determine the ground-state energy \cite{Katsura-PRL13}. The 
dimension of the local Hilbert space basis is fixed by choosing a maximum occupation number $\hat{n}_{max}$. 
We chose $\hat{n}_{max}= 5$ in order to guarantee accurate results, keeping up to $m=300$ states per block, and 
obtained a discarded weight around $10^{-5}$ or less.  We set our energy scale choosing $V_0=1$ and 
explored the phase diagram by changing the parameter $V_2$.\par 
\begin{figure}[t]
\begin{minipage}{17.5pc}
\includegraphics[width=17.5pc]{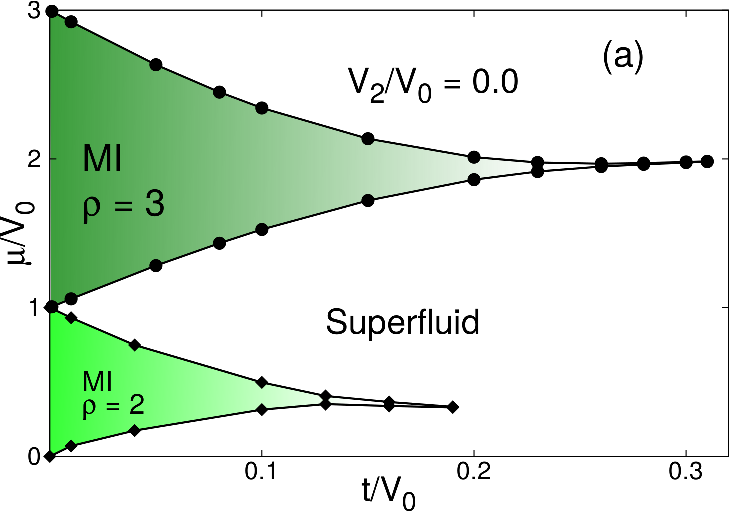}
\end{minipage}
\hspace{2pc}%
\begin{minipage}{17.5pc}
\includegraphics[width=17.5pc]{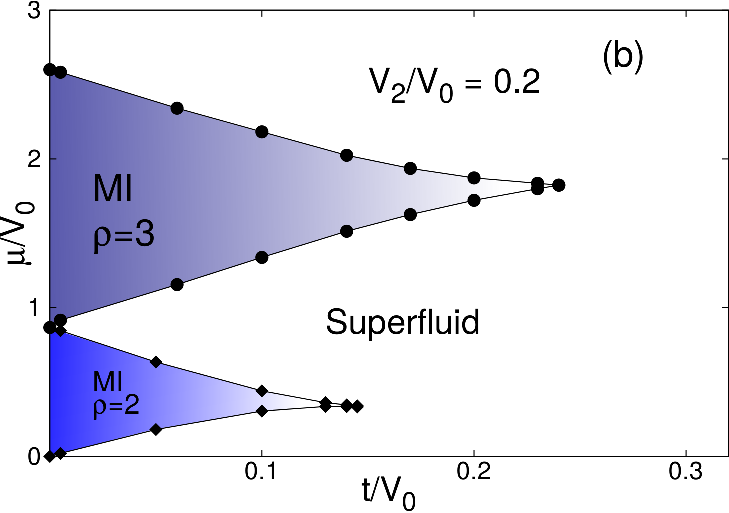}
\end{minipage}
\hspace{2pc}%
\begin{minipage}{17.5pc}
\includegraphics[width=17.5pc]{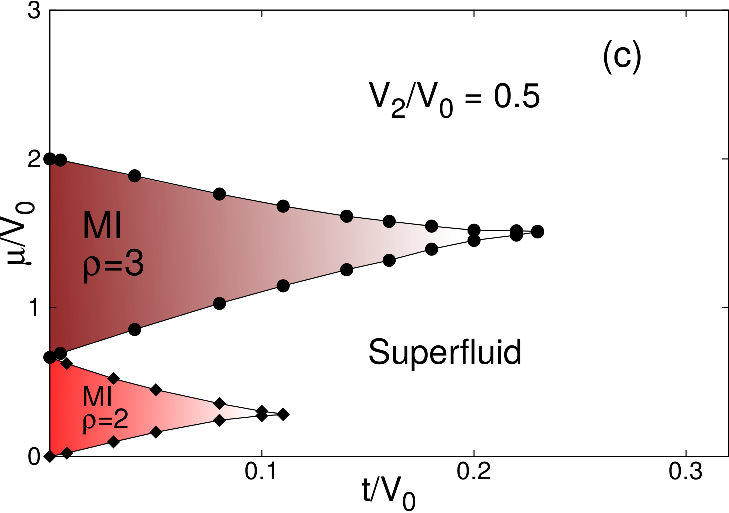}
\caption{\label{fig3}   
Phase diagram for the first two Mott lobes of the  spin-1 Bose-Hubbard model with 
three-body interaction and antiferromagnetic spin interaction. (a) $V_2/V_0=0$, (b) $V_2/V_0=0.2$,  and (c) $V_2/V_0=0.5$. The solid lines are 
visual guides.}
\end{minipage}
\hspace{2pc}%
\end{figure}
\subsection{Antiferromagnetic case ($V_2>0$)}
First of all, we analyze the system in the non-hopping case $t \rightarrow 0$, the atomic limit. For $n_i$ 
particles, the energy is given by
\begin{eqnarray}
E_M(n_i)& = & \frac{V_0}{6}n_i(n_i-1)(n_i-2)\nonumber \\
        &  & + \frac{V_2}{6}\left[<\mathbf{\hat{S}^2_i}> - 2n_i\right](n_i-2).\nonumber
\end{eqnarray}
For the antiferromagnetic case, we need to separate the analysis for even and odd densities. The 
expectation value $<\mathbf{\hat{S_i}^2}>=S(S+1)$ has an important role at this point: if the system has an even 
filling, it could have a ground state composed of singlets at each site ($S=0 \rightarrow <\mathbf{\hat{S_i}^2}>=0$), whereas for 
odd filling there will always be an unpaired boson at each site ($S=1 \rightarrow <\mathbf{\hat{S_i}^2}>=2$). So for even 
filling, $n_e$, the energy of the Mott lobes is $E_M(n_e)= V_0n_e(n_e-1)(n_e-2)/6-V_2n_e(n_e-2)/3 $, while for 
odd filling, $n_o$, $E_M(n_o)= V_0n_o(n_o-1)(n_o-2)/6+V_2(1-n_o)(n_o-2)/3 $. Thus the boundaries of the 
Mott lobes can be determined from the energy difference $\mu_0(n\rightarrow n+1)=E_M(n+1)-E_M(n)$ between the Mott 
ground state with $n$ and with $n+1$ particles (here $\mu_0$ represents the chemical potential at $t=0$). Taking into account the above definition, 
we obtain $\mu_0(n_e\rightarrow n_{e+1})=n_e(n_e-1)V_0/2-n_eV_2/3$ and $\mu_0(n_o\rightarrow n_{o+1})=n_o(n_o-1)V_0/2-(n_o-1)V_2$.\par  
We can see that the borders of the Mott lobes decrease due to the spin interaction, regardless of the global density, in 
contrast with the model with two-body interactions, for which the upper border of the second Mott lobe remains fixed at $\mu_0/U_0=2$.
Therefore, for spinor bosons interacting with three-body terms, we obtain that at the atomic limit the upper border of the first (second ) 
Mott lobe is given by $\mu^*_0/V_0=1-\frac{2V_2}{3V_0}$ ($\mu^*_0/V_0=3-2\frac{V_2}{V_0}$).\par
The behavior of the chemical potential with respect to the system size is shown in Fig. \ref{fig1} for global density $\rho=N/L=2$ 
and $V_2/V_0=0.2$. In the left panel (Fig. \ref{fig1}a), we consider $t/V_0=0.1$ and see that the 
energy for adding (removing) a particle decreases (increases), following a quadratic behavior, as the lattice 
size increases. It is possible to determine that the value of the energy gap at the thermodynamic 
limit ($N,L \rightarrow \infty $) is finite and not zero $\Delta/V_0 = 0.136$, and due to the integer number 
of atoms at each lattice site, the system is in a Mott insulator state. Increasing the hopping parameter to $t/V_0=0.145 $, 
we observe that the energy for adding (removing) a particle decreases (increases), following a 
linear behavior, as the lattice size increases, and at the thermodynamic limit the value is the same, i. e. 
the gap vanishes and the system is in a superfluid state (Fig. \ref{fig1}b). We stated based on  Fig. \ref{fig1} that the system undergoes a quantum 
phase transition as the kinetic energy of the bosons increases.\par  
\begin{figure}[t]
\begin{minipage}{19pc}
\includegraphics[width=19pc]{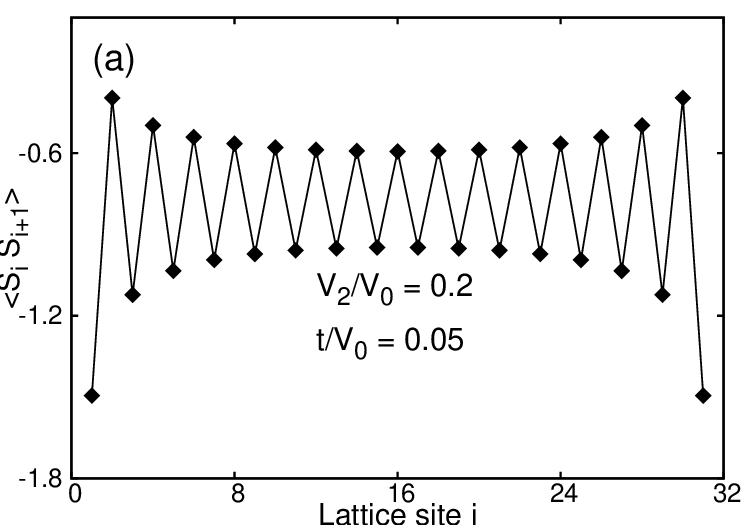}
\end{minipage}
\hspace{5pc}%
\begin{minipage}{19pc}
\includegraphics[width=17pc]{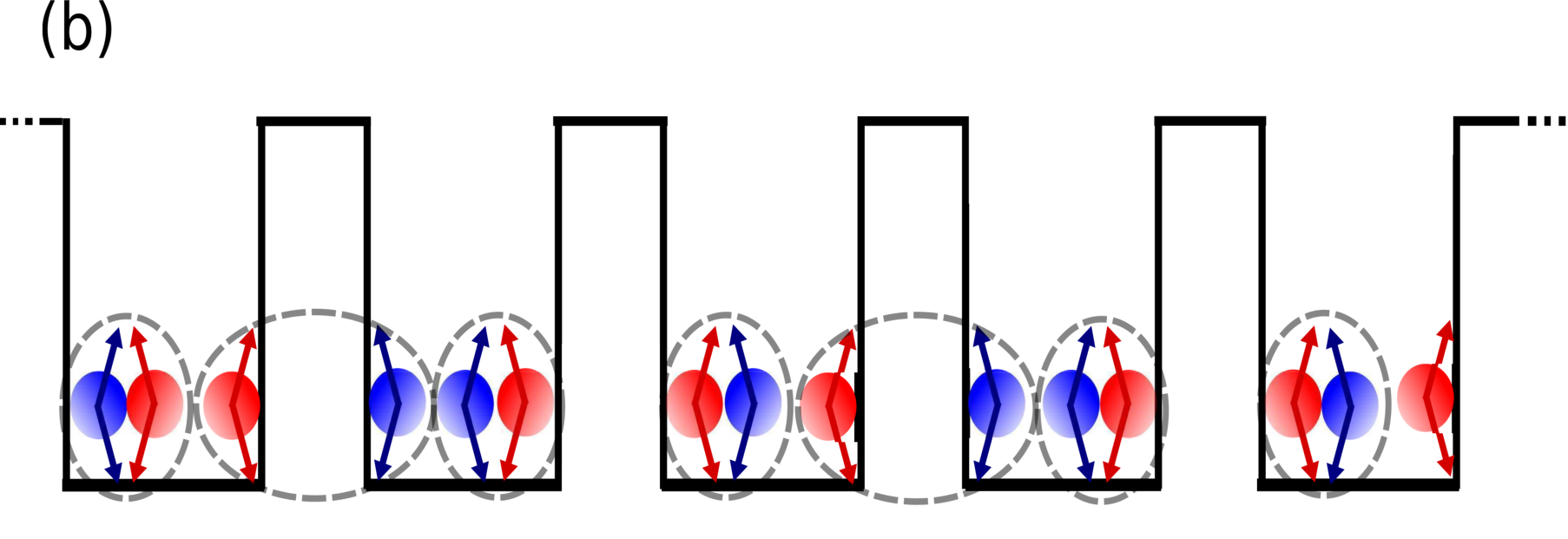}
\caption{\label{fig4} (a) Spin-spin correlations for a spinor boson chain with antiferromagnetic spin-dependent interaction. The
parameters used are $\rho=3$, $L=32$, and $t/V_0=0.05$, which correspond to the second Mott insulator lobe. (b) Schematic of a possible microscopic 
configuration for the dimerized Mott state for $\rho=3$.}
\end{minipage}\hspace{2pc}%
\end{figure}
Varying the number of particles and calculating the ground-state energy for each set of parameters with a fixed lattice size, we can determine the 
behavior of the chemical potential as a function of the global density, which is shown is Fig. \ref{fig2}a for a spinor chain with antiferromagnetic 
spin-dependent interaction $V_2/V_0=0.2$ and a hopping parameter $t/V_0=0.05$. We observe that as the global density increases, the chemical potential 
increases monotonously, passing through the value $\rho=1$  without major changes, which indicates that the Hamiltonian (\ref{Hgen}) does not exhibit 
a Mott insulator state for a filling of one atom per site, a result that is well known for the model with two-body interactions. This happens because the 
quantum fluctuations are not large enough to generate a Mott insulator state with three-body local interactions. The continuous growth of the 
chemical potential as a function of the global density ceases when the density reaches integer values, i. e. for $\rho=2$ and $\rho=3$ we observe 
a plateau. Note that the second plateau is larger than the first, which reflects the major localization of the particles for this filling, a fact 
that was previously reported for spinless bosons with three-body interaction terms.\par 
An important fact is that the compressibility $\kappa=\partial \rho /\partial \mu$ is positive (see Fig. \ref{fig2}a). An opposite result means that 
the  transitions are of the first order kind, which was stated for two-dimensional systems by Batrouni {\it et al.}~\cite{Batrouni-PRL00}  and  
de Forges de Parny~\cite{Forges-PRB13} for spinless and spin-1 bosons, respectively. But for a spinor boson chain with two-body interactions, it was 
shown that transitions for even Mott lobes are of the first order, despite the fact that the compressibility is positive, indicating that the compressibility 
criterion is not sufficient to determine the type of transition~\cite{Batrouni-PRL09}. Taking into account the above, we calculate the spin population 
fractions $N_+/N=N_-/N$ and $N_0/N$ as a function of the global density (Fig. \ref{fig2}b). There are two different regions in this figure; for  
$\rho\leq2$, the dominant projection is $\sigma=0$ and $N_0/N >> N_+/N$. This implies that ground state for $\rho=2$ is not composed of singlets, 
because if so we would have to obtain that $N_+/N=N_-/N=N_0/N$, and this clearly does not happen. In the inset of Fig. \ref{fig2}a, we show that the square  
of the local moment ($<S^2>$) as a function of the hopping parameter $t/V_0$ for a fixed global density $\rho=2$ remains around the same value 
$<S^2>\sim 3.2$ as the hopping increases from zero. This means that inside of the Mott insulator region (left of the vertical dashed line) $<S^2>$ is 
finite and non-zero, which means that the ground state is not composed of singlets, because if so, $<S ^2>\rightarrow 0$ as $t/V_0$ goes to zero.\par 
Based on Fig. \ref{fig2}b, we see that inside of the superfluid region, the global density drives a change in the ground state of the system, i. e. a 
quantum phase transition takes place 
around $\rho\approx2.2$, and the system passes form a state where $N_0/N >> N_+/N$ to a state with $N_+/N > N_0/N$, a fact that is a consequence of 
spin-dependent three-body interactions. Also, we observe in Fig. \ref{fig2}b that around integer densities the ground state is the same, which 
indicates that the quantum phase transitions will be of the first order for both even and odd Mott lobes. This result diverges from the two-body 
findings, where for even and odd Mott lobes first and second order transitions were found, respectively~\cite{Batrouni-PRL09}.\par 
Knowing that the ground state of our system can be gapped or gapless, we can use this fact to draw a phase diagram, which is shown in 
Fig. \ref{fig3} for the first two Mott lobes ($\rho=2$ and $\rho=3$) in 
the $(\mu/V_0,t/V_0)$ plane, for different values of spin dependent interaction $V_2$. We observe that the Mott lobes 
are surrounded by a superfluid phase, similar to the spinor case with two-body interactions~\cite{Rizzi-PRL05}.
In Fig. \ref{fig3}a, we note that the phase diagram for $V_2/V_0=0$ coincides with the phase diagram 
predicted and reported for the spinless case with three-body interactions~\cite{JSV-EPJB12, Sowinski-PRA12}. 
Now the spin-dependent interaction is turned on, and we show in Fig. \ref{fig3}b and Fig. \ref{fig3}c that the Mott 
lobes are suppressed as the spin-dependent interaction parameter increases, implying that the position of the 
critical point moves to the left (small values of $t/V_0$). This behavior contrasts with the results for the 
two-body case, where the even Mott lobes grow while the odd ones decrease, exhibiting an even-odd asymmetry. 
Note that in Fig. \ref{fig3} the numerical results at $t=0$ are in accord with the atomic limit results.\par  
\begin{figure}[t]
\includegraphics[width=19pc]{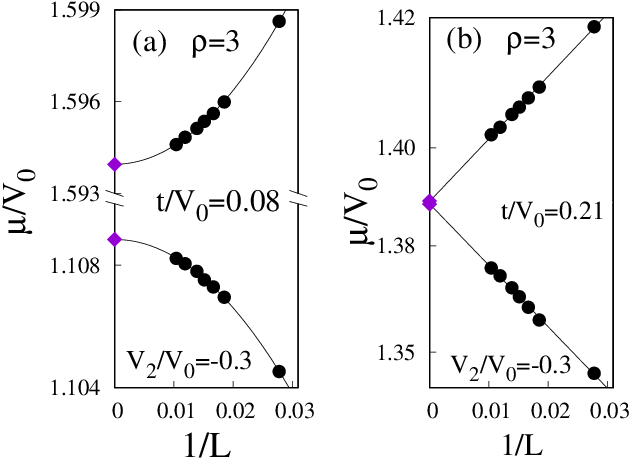}
\caption{\label{fig5} Chemical potential as a function of the inverse of the lattice size for $\rho=3$ and $V_2/V_0=-0.3$. 
The left panel shows a state with a finite gap for $t/V_0=0.08$. The right panel shows a gapless behavior for $t/V_0=0.21$.
The values at the thermodynamic limit (diamonds) were obtained by fitting the DMRG results (circles) in the left (right) panel to the function 
$\mu^{p,h}(L)=\tilde{\mu}^{p,h}+ c_1/L + c_2/L^2$  ($\mu^{p,h}(L)=\tilde{\mu}^{p,h}+ c_3/L$), where $c_i$ are constants.
The lines are visual guides.}
\end{figure}
According to several authors, for the spin-1 Bose-Hubbard model with two-body interactions, the Mott lobes exhibit 
interesting magnetic properties: some of these lobes can be in a dimerized, singlet, or nematic phase. For instance, in order to observe the dimerization, 
the spin-spin correlation function, the dimer susceptibility, or the dimer order parameter has been calculated in various 
studies~\cite{Chen-PRA12,Apaja-PRA06,Rizzi-PRL05}. In Fig. \ref{fig4}a, we show the spin-spin correlation function between neighboring sites
for the second Mott insulator lobe ($\rho=3$), and the parameters considered are $L=32$, $t/V_0=0.05$, and $V_2/V_0=0.2$. 
Disregarding the border effects due to the open boundary conditions, we observe that 
spin-spin correlations between neighboring sites are ferromagnetic and exhibit an oscillating behavior around two values, which reflects the fact that 
there is a unit cell composed of two sites that repeats on the lattice; hence the odd Mott lobes have a dimer magnetic order, in a manner similar 
to the model with two-body interactions~\cite{Yip-PRL03,Bergkvist-PRA06,Rizzi-PRL05,Apaja-PRA06}. In Fig. \ref{fig4}b, we present a possible 
dimerized configuration following F. Zhou's illustrations~\cite{Zhou-EL03}. Here, the wells represent lattice sites, the balls represent the atoms, 
and three bosons per site were considered. Each boson has three spin orientations, of which we only show two (blue and red balls). Each pair of 
blue and red balls forms a singlet, and we can see a possible way in which the translational symmetry is broken. The dashed lines explicitly show the 
singlets that can form along the lattice, and we can observe that an alternating atom-atom correlation or singlet-singlet correlation between 
neighboring sites can happen, which explains the spin correlation function obtained in Fig. \ref{fig4}a.\par 

\begin{figure}[t]
\begin{minipage}{17.5pc}
\includegraphics[width=17.5pc]{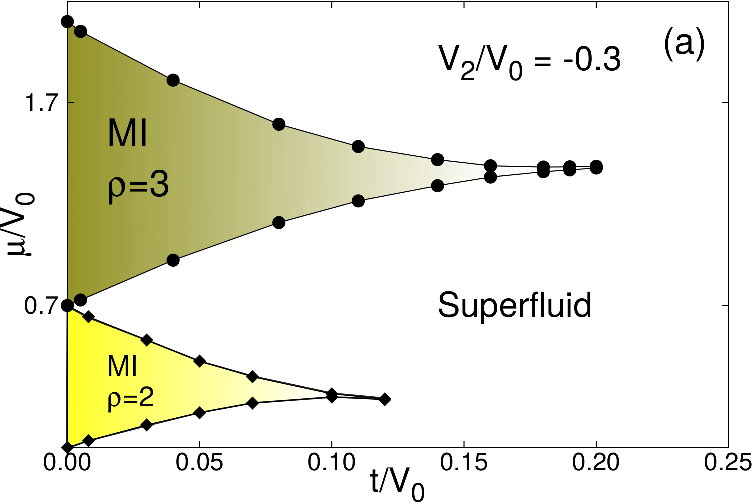}
\end{minipage}
\hspace{2pc}%
\begin{minipage}{17.5pc}
\includegraphics[width=17.5pc]{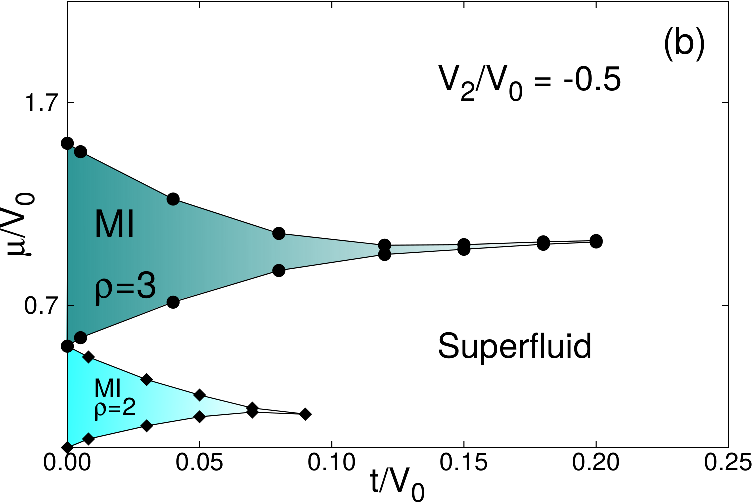}
\end{minipage}
\hspace{2pc}%
\begin{minipage}{17.5pc}
\includegraphics[width=17.5pc]{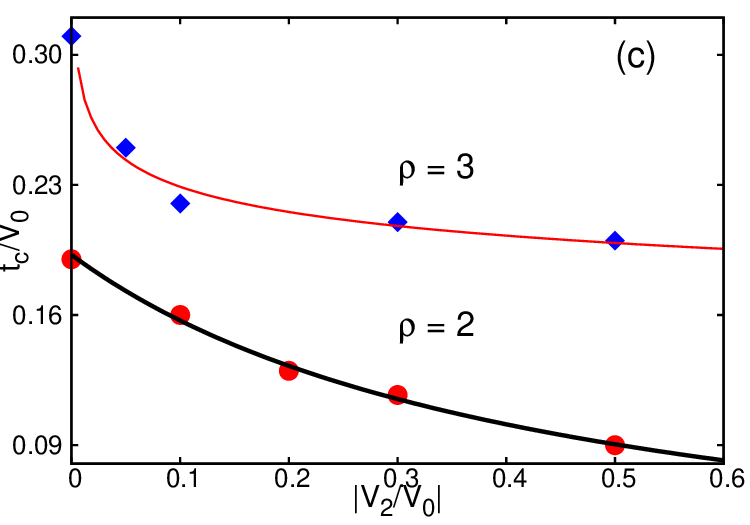}
\caption{\label{fig6}   
Phase diagram for the first two Mott lobes of the $S=1$ Bose-Hubbard model with 
three-body interaction and ferromagnetic spin interaction. The panels correspond to different values 
of $V_2/V_0=-0.3$ (a) and $V_2/V_0=-0.5$ (b). The solid lines are visual guides. (c) Evolution of the critical points with the spin exchange 
interaction parameter $|V_2/V_0|$. The solid lines correspond to the fit to the expressions (\ref{eq:2}) and (\ref{eq:3}) for the lobes $\rho=2 $ and 
$\rho=3 $, respectively.}
\end{minipage}
\hspace{2pc}%
\end{figure}
\subsection{Ferromagnetic case ($V_2<0$)}

Now we consider a lattice of spin-1 atoms with a ferromagnetic local spin-dependent interaction.
At the atomic limit, we expected that the border among the Mott lobes would depend on the spin-dependent strength, and to find the precise dependence, we 
have to take into account that the local spin must be at a maximum; hence $<\mathbf{\hat{S}^2_i}>=n_i(n_i+1)$ for all Mott lobes, where $n_i$ is 
the number of atoms per site. Considering the above fact, the ground-state energy with $n_i$ particles is 
$E_M(n_i)=\frac{V_0+V_2}{6}n_i(n_i-1)(n_i-2)$. Therefore, the border between Mott lobes with $n_i$ and $n_i+1$ particles is given by 
$\mu_0(n_i\rightarrow n_i+1)= \frac{V_0+V_2}{2}n_i(n_i-1)$. Then the upper border of the Mott insulator lobes depends on the local spin-dependent 
interaction, and we obtain $\mu^*_0/V_0=1+\frac{V_2}{V_0}$ and $\mu^*_0/V_0=3+3\frac{V_2}{V_0}$ for the first and the second Mott lobe, 
respectively. Note that the above atomic limit expressions are in accordance with the numerical results shown in Figs. \ref{fig6}a and 
\ref{fig6}b.\par
In Fig. \ref{fig5}, we show the evolution of the chemical potential with the lattice size, fixing the global density at $ \rho=3$ and $V_2/V_0=-0.3$. 
If we consider the kinetic energy parameter to be less than the repulsion interaction ($t/V_0=0.08$), we expect that at the thermodynamic limit, the chemical 
potential to increase or decrease the number of particles in one will tend to different values, and so a finite energy gap of $\Delta/V_0 \approx 0.485$
will be obtained (see Fig. \ref{fig5}a). Note that the above gapped state is obtained for an integer density and is due to the interaction between the atoms; 
therefore this is a Mott insulator state, for which we also observe that the energy for adding (removing) a particle decreases (increases) following a
quadratic behavior, regardless of the sign of the local spin-dependent interaction. The increase of the kinetic energy ($t/V_0=0.21$) leads to more 
quantum fluctuations. The atoms tend to delocalize throughout the lattice, and the energy for adding or removing a particle follows a linear behavior with the 
inverse of the system size, reaching the same value at the thermodynamic limit. Therefore, for these parameter values the ground state is gapless and 
corresponds to a superfluid state (see Fig. \ref{fig5}b).\par 
\begin{figure}[t]
\begin{minipage}{17.5pc}
\includegraphics[width=17.5pc]{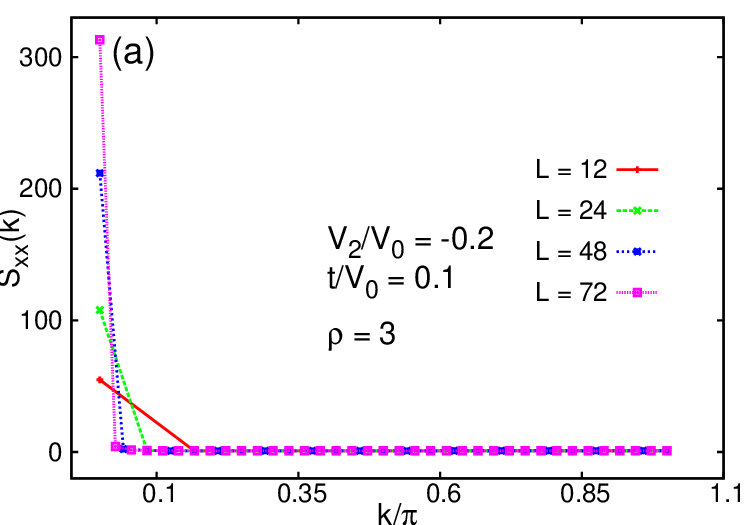}
\end{minipage}
\hspace{2pc}%
\begin{minipage}{19pc}
\includegraphics[width=17pc]{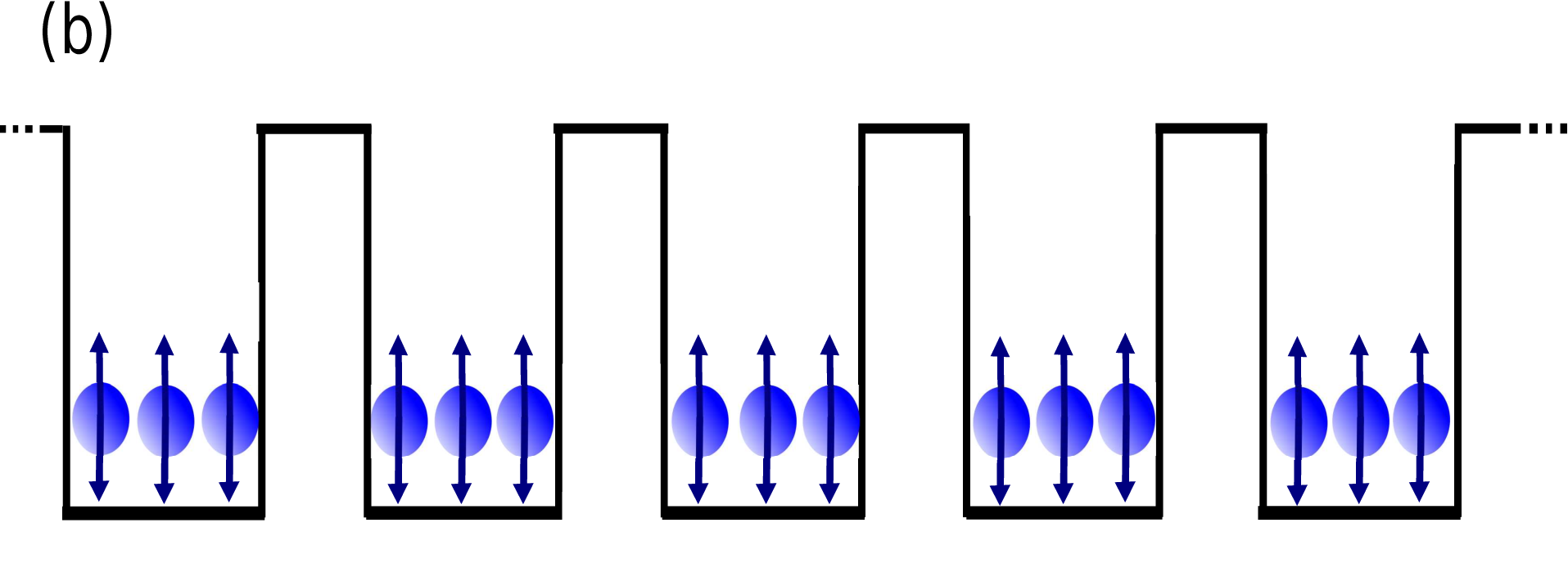}
\end{minipage}
\hspace{2pc}%
\begin{minipage}{17.5pc}
\includegraphics[width=17.5pc]{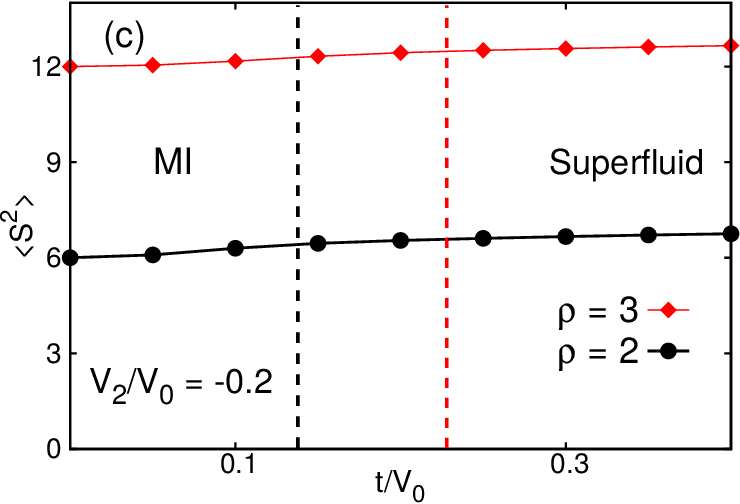}
\caption{\label{fig7} (a) The $x$-axis spin structure factor $S_{xx}(k)$ for a spinor boson chain with ferromagnetic spin-dependent 
interaction inside of the second Mott lobe ($\rho=3$). Here  $V_2/V_0=-0.2$ and $t/V_0=0.1$. (b) Schematic of a microscopic configuration for 
``long-range'' ferromagnetic order. (c) The expectation value $<\mathbf{\hat{S}^2_i}>$ as a function of the hopping $t/V_0$ for the two first lobes with 
$V_2/V_0=-0.2$.}
\end{minipage}
\hspace{2pc}%
\end{figure}
The phase diagram of spin-1 bosons with ferromagnetic local spin-dependent interaction in the plane ($\mu/V_0$,$t/V_0$) is shown Fig. \ref{fig6}. It 
is composed of a superfluid phase that surround the Mott insulator lobes, which decrease when the absolute value of the three-body spin-dependent 
strength increases. In Fig. \ref{fig6}a and Fig. \ref{fig6}b, we observe that for a fixed number of bosons, the critical point that 
separates the Mott insulator and the superfluid regions evolves in a different way, depending on the global density:  the first (second) Mott lobe 
decreases quickly (slowly). Using the vanishing gap criteria, we estimate the position of the critical point for each value of $V_2$ and obtain the
points of Fig. \ref{fig6}c, where it is possible to see  the differing  behavior of the critical points of the first and second Mott lobes. 
The lower solid line shown in this figure is the adjusted fit for $\rho=2$ using the expression  
\begin{equation}\label{eq:2}
\frac{t_c}{V_0}=\frac{1}{a\frac{V_2}{V_0}+b},
\end{equation}
\noindent where $a=11.63\pm0.51$ and $b=5.20\pm0.17$ are constants, while for $\rho=3$, we describe the evolution of the critical point following 
the expression
\begin{equation}\label{eq:3}
\frac{t_c}{V_0}=\alpha\left(\frac{V_2}{V_0}\right)^\beta+\gamma,
\end{equation}
\noindent where the constants take the following values: $\alpha=0.115\pm0.081$, $\beta=0.166\pm0.071$, and $\gamma=0.304\pm0.035$.\par 
Also, in the second Mott lobe it is possible to observe a reentrant behavior; however, we believe that this is due to the truncation of the local 
Hilbert space, and therefore if we increase the number of possible states in the local Hilbert space, this may not be observed.\par 
In order to determine the magnetic order and to determine how the local moments are oriented from site to site, we calculated the spin structure factor 
by means of the correlation functions given by
\begin{equation}
\label{eqsf}
S_{\sigma \sigma} (k) = \sum_{\ell} e^{i k \ell}\left< S_{\sigma,j+\ell}S_{\sigma,j}\right>,
\end{equation}
\noindent where $\sigma$ can be $x$ or $z$.\par  
In Fig.~\ref{fig7}a, we show $S_{xx}(k)$ in the Mott insulator state at $t/V_0=0.1$. We consider three 
atoms per lattice site and ferromagnetic interaction $V_2/V_0=-0.2$.  We obtain a peak at $k=0$, indicating 
that the Mott phase has a ferromagnetic order. We can see that this peak grows with the lattice size, 
showing a ``long-range'' ferromagnetic order. A schematic of the microscopic configuration of the atoms is shown in Fig.~\ref{fig7}b,  where it is 
possible to observe all the local moments in the same direction (we chose the ``up'' direction to illustrate the behavior).
During the discussion of the atomic limit, we said that to consider a ferromagnetic interaction leads to the 
expectation value $<\mathbf{\hat{S}^2_i}>$ being a maximum. This is shown in Fig. \ref{fig7}c, where we 
calculate the value of $<S_i^2>$ as a function of $t/V_0$ and observe that this quantity remains constant for $\rho=2$ and $3$. Note that the 
numerical value corresponds to   $<S_i^2 >= n_i(n_i + 1 ) \approx 6$ and $12$, respectively. Also, we calculate the magnetic structure factor in 
the superfluid phase, and it is also ferromagnetic.

\section{\label{sec4} Conclusions}

In this paper, we studied the ground state of spin-1 bosons loaded in a one-dimensional optical 
lattice, using the $S=1$ Bose-Hubbard model with local three-body interaction terms at zero temperature. 
In order to determine the phase diagram, we employed the density matrix renormalization group method with a maximum of five bosons per site, and we 
obtained that the ground state of the model exhibits a superfluid or a Mott insulator state, regardless of the sign of the local spin-dependent interaction.
The Mott insulator region with one boson per site is absent in the phase diagrams, while for larger densities the Mott lobes increase with the density, 
reflecting a higher localization in the system. We show that at the atomic limit the borders between different Mott lobes always depend on the local 
three-body spin-dependent interaction, a fact that leads to the Mott insulator regions always decrease as the spin-dependent strength parameter 
increases for both ferromagnetic and antiferromagnetic interaction. The above statement contains an important result, which is that for 
antiferromagnetic interaction, a one-dimensional system of spin-1 bosons that interact under local three-body interactions, the even-odd asymmetry is 
absent, which differs from the two-body case, where this asymmetry is the most relevant result, remarked by several authors.\par 
For antiferromagnetic spin-dependent interaction, we observe that the odd lobes are dimerized, while the even lobes are not composed of singlets. 
When the global density increases while the other system parameters are fixed, the spinor chain passes from a superfluid to a Mott insulator state for 
densities larger than one. These quantum phase transitions are of the first-order kind for both even and odd lobes. Also, the spin-dependent interaction 
generates a quantum phase transition inside of the superfluid region for a density $\rho\approx2.2$, regardless of the hopping parameter value.\par 
For ferromagnetic coupling, both the Mott insulator and the superfluid phase exhibit a ``long-range'' ferromagnetic order, and we found that the critical points 
move to lower values as the spin interaction increases, and their evolution depends on the global density.\par

\section*{Acknowledgments}
The authors are thankful for the support of DIEB- Universidad Nacional de Colombia and Departamento Administrativo de Ciencia, Tecnolog\'{\i}a e 
Innovaci\'on (COLCIENCIAS) (grant No. FP44842-057-2015). J.S.-V. and R.F. are grateful for the hospitality of the ICTP, where part of this work 
was done.

\bibliography{Bibliografia}

\end{document}